\documentclass[superscriptaddress,twocolumn,10pt]{revtex4-1}
\usepackage{amsmath}
\usepackage{graphicx}
\usepackage{verbatim}
\usepackage{color}
\usepackage{subfigure}
\usepackage{hyperref}
\usepackage{amssymb}
\usepackage{amsthm}
\usepackage{amsfonts}

\begin{document}

\title{Probing Patchy Saturation of Fluids in Nanoporous
Media by Ultrasound}
\author{Boris Gurevich}
\affiliation{Curtin University, Perth, Western Australia, Australia} %
\affiliation{CSIRO, Perth, Western Australia, Australia.}
\author{Michel M. Nzikou}
\affiliation{Curtin University, Perth, Western Australia, Australia} %
\author{Gennady Y. Gor}
\email[Corresponding author, e-mail: ]{gor@njit.edu}
\affiliation{Department of Chemical, Biological and Pharmaceutical
Engineering\\ New Jersey Institute of Technology, Newark, NJ, USA}
\date{\today}

\begin{abstract}
Nanoporous materials provide high surface area per unit mass and are capable of fluids adsorption. While the measurements of overall amount of fluid adsorbed by a nanopororus sample are straightforward, probing the fluid spacial distribution is non-trivial. We consider published data on adsorption and desorption of fluids in nanoporous glasses reported along with the measurements of ultrasonic waves propagation. We analyse these using Biot's theory of dynamic poroelasticity, approximating the patches as spherical shells. Our calculations show that on adsorption the patch diameter is on the order of 10-20 pore diameters, while on desorption the patch size is comparable to the sample size. Our analysis suggests that one can employ ultrasound to probe the uniformity of fluid spatial distribution in nanoporous materials.
\end{abstract}

\maketitle

\vspace{2cm}

Many natural and synthetic materials of industrial relevance have nanoporous structure, providing high surface area per unit mass and being capable of fluids adsorption. While the overall amount of fluid adsorbed by a nanopororus sample can be routinely measured \cite{Rouquerol2013}, the spatial distribution of fluid inside a nanoporous sample is not easy to probe. Yet, the non-uniformity of the fluid distribution in a nanoporous sample affects many of its physical properties. Adsorption-induced stresses strongly depend on the saturation of pores \cite{Gor2017review}, therefore spatial distribution of fluid affects the strains in nanoporous materials. The strains in its own turn can affect the permeability of nanoporous media \cite{Pan2007}. Another example is the change of optical properties of porous glasses during fluids adsorption \cite{Barthelemy2007, Varanakkottu2014}. Thus there is a clear demand to extract the information about fluid distribution from experimental measurements. 

If a sample of a mesoporous material is placed in vapor at a pressure below the saturation pressure, some of the vapor is adsorbed on the pore walls. The amount of condensate adsorbed on the pore walls increases when the vapor pressure is increased (adsorption process) and decreases when the pressure is reduced (desorption process). However, it is believed that during adsorption and desorption processes, distributions of the confined fluid and vapor in the pores are very different. On adsorption, the thickness of the condensate film increases steadily and uniformly in all pores, whereas on desorption the fluids form macroscopic patches \cite{Page1995}. Indeed optical techniques show formation of macroscopic patches during desorption \cite{Page1995}. No such behavior has been shown for adsorption; yet the fluid distribution during the adsorption processes has not been studied in detail.

\begin{figure}[ht!]
\centerline{\includegraphics[width=0.7\linewidth]{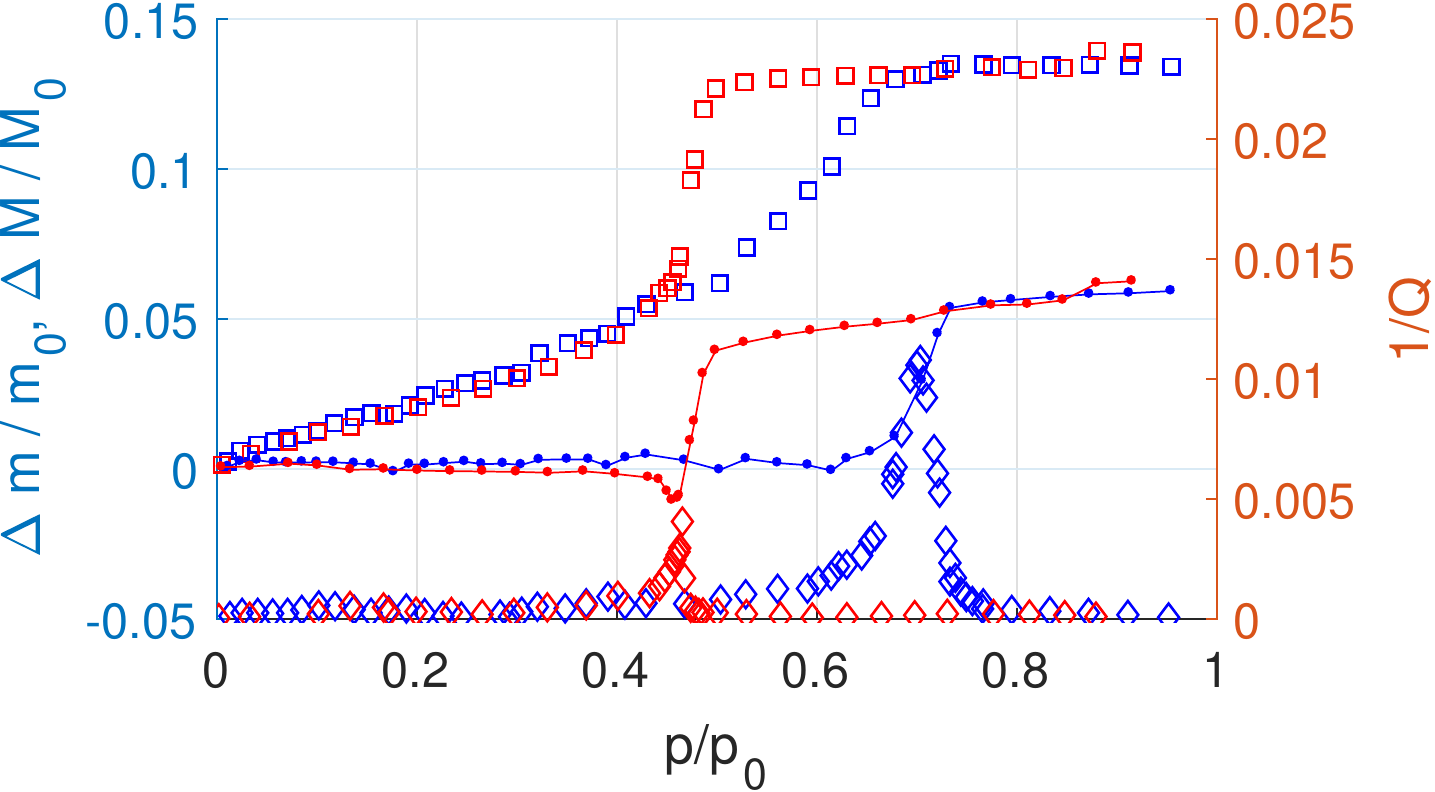}}
\caption{Relative change in the mass $\Delta m/m$ (squares), in the longitudinal modulus $\Delta M/M$ (lines) and attenuation factor $1/Q$ (diamonds) during adsorption (blue) and desorption (red) of n-hexane in Vycor\cite{Page1995}.}
\label{fig:hexane_data}
\end{figure}

\begin{figure}[ht!]
\centerline{\includegraphics[width=0.7\linewidth]{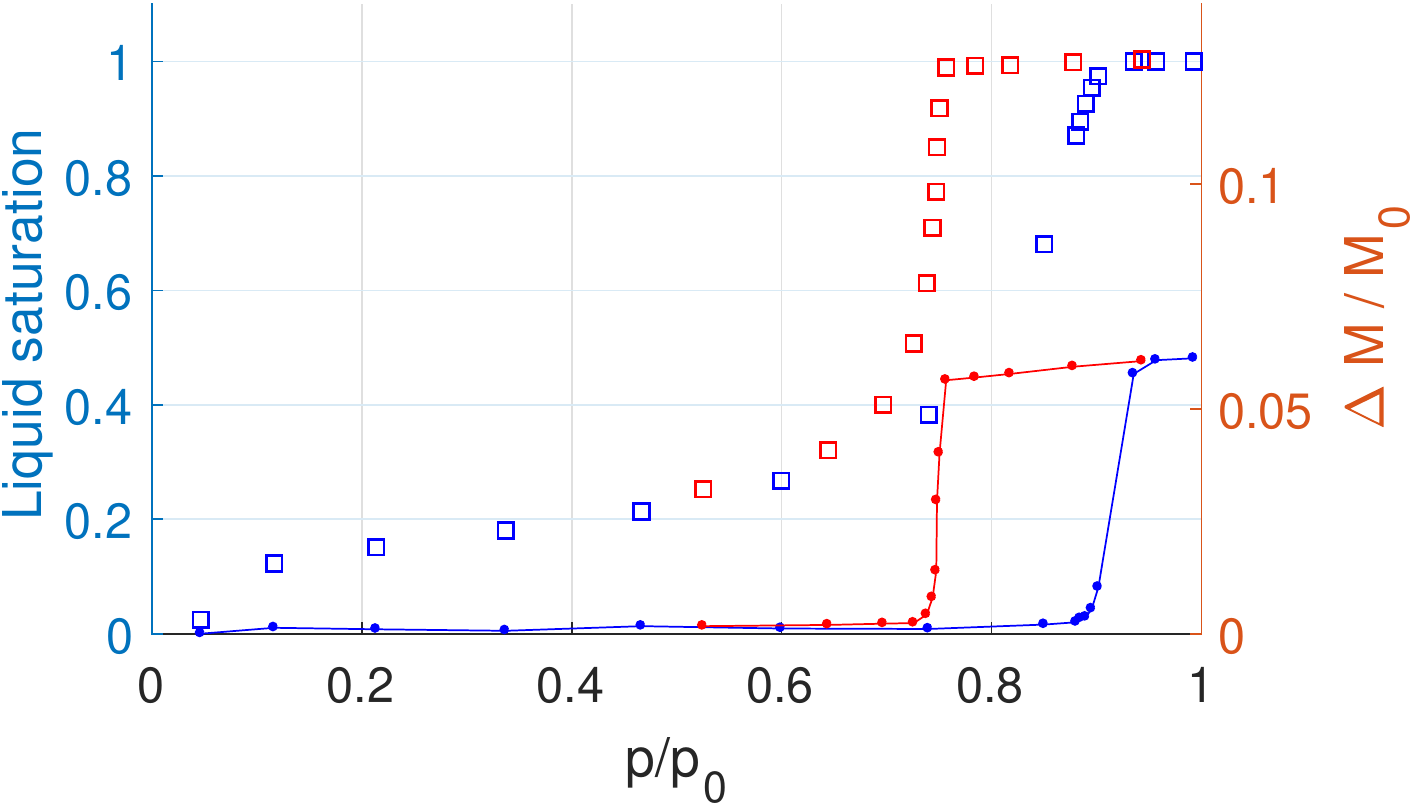}}
\caption{Mass fraction of liquid argon (squares) and relative change in the longitudinal modulus $\Delta M/M$ (lines) during adsorption (blue) and desorption (red) of argon in Vycor\cite{Schappert2014}.}
\label{fig:argon_data}
\end{figure}

One method that can shed light on details of fluid distribution during sorption (adsorption and desorption) is the ultrasonic technique, and specifically dependence of ultrasonic velocity on saturation of the pore space. The theory of poroelasticity shows that the dependence of elastic modulus on liquid saturation is controlled by the spatial distribution of fluids \cite{Toms:etal:2007}. Indeed, ultrasonic data \cite{Page1995,Schappert2014} show that the increase of the longitudinal modulus of the nano-porous glass near the capillary condensation point is sharp but not instant, see Fig.\ \ref{fig:hexane_data} and  \ref{fig:argon_data}. This suggests that it could be possible to analyze the fluid distribution in sorption experiments using the dependence of elastic properties on liquid saturation obtained from ultrasonic data.

The analysis of patchy saturation of porous media is a topical issue in petroleum geophysics (see Refs. \onlinecite{Domenico:1976,Dutta:Ode:1979a,Dutta:Ode:1979b,Murphy1982,Cadoret:etal:1998,Knight:Dvorkin:Nur:1998,Johnson:2001,Toms:etal:2007, Toms2008,Caspari:Muller:Gurevich:2011,Rubino:Holliger:2012} and references therein). However, application of dynamic saturation models to the data measured on rocks is often problematic, because the patchy saturation effects on the velocity and amplitude of ultrasonic waves in such complex porous media are often obscured by other phenomena, such as squirt flow \cite{Mavko:Nur:1975,Jones:1986,Murphy:etal:1986,Mavko1991,Gurevich2010,Muller:Gurevich:Lebedev:2010}. Nanoporous Vycor glass, which has narrow pore size distribution and uniform mechanical properties provides an excellent medium for testing those models. Furthermore, adsorption processes in such uniform media result in extremely uniform saturation of the pore space, which is impossible to achieve in natural materals. 

We consider two experimental works, reporting ultrasonic measurements during vapor adsorption on nanoporous Vycor glass:
adsorption of n-hexane at room temperature \cite{Page1995} and of argon at cryogenic temperature \cite{Schappert2014}. The longitudinal moduli of the samples obtained from the velocity of ultrasonic waves as a function of vapor pressure are shown in Fig.\ \ref{fig:hexane_data} (for n-hexane) and \ref{fig:argon_data} (for argon) along with the saturation. Figure \ref{fig:hexane_data} also shows the data on wave attenuation from Ref. \onlinecite{Page1995}.

Bulk modulus of a porous medium with porosity $\phi$ and bulk modulus of the empty matrix $K_{0}$, made up of a solid with bulk modulus $K_{s}$ and saturated with a single fluid with a bulk modulus $K_{f}$ is given by the Gassmann equation \cite{Gassmann1951,Berryman1999}, application of which has been recently demonstrated for nanoporous media \cite{Gor:Gurevich:2017}
\begin{equation}
\label{Gassmann}
K_{G}(K_{f})=K_{0}+\frac{\alpha ^{2}}{\frac{\alpha -\phi }{K_{s}}+\frac{\phi}{K_{f}}},
\end{equation}
where $\alpha =1-K_{0}/K_{s}$ is the Biot-Willis coefficient. For weak fluids (relative to the solid matrix), $K_{f}\ll K_{s},$ equation \ref{Gassmann} can be linearized in $K_{f}$ to give
\begin{equation}
\label{Gassmann_linear}
K_{G}(K_{f}) \simeq K_{0}+\frac{\alpha^2}{\phi }K_{f}. 
\end{equation}

\begin{figure}[ht!]
\centerline{\includegraphics[width=0.7\linewidth]{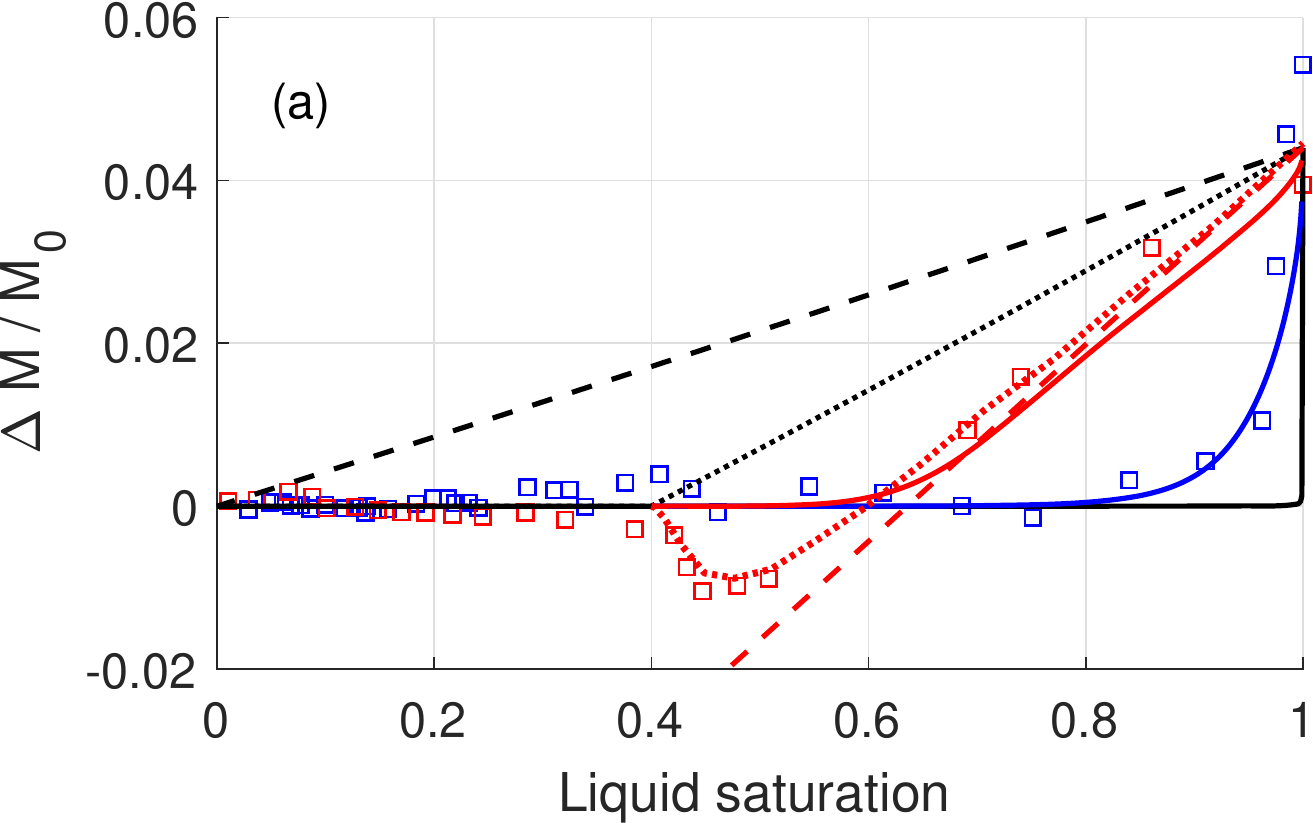}} \centerline{\includegraphics[width=0.7\linewidth]{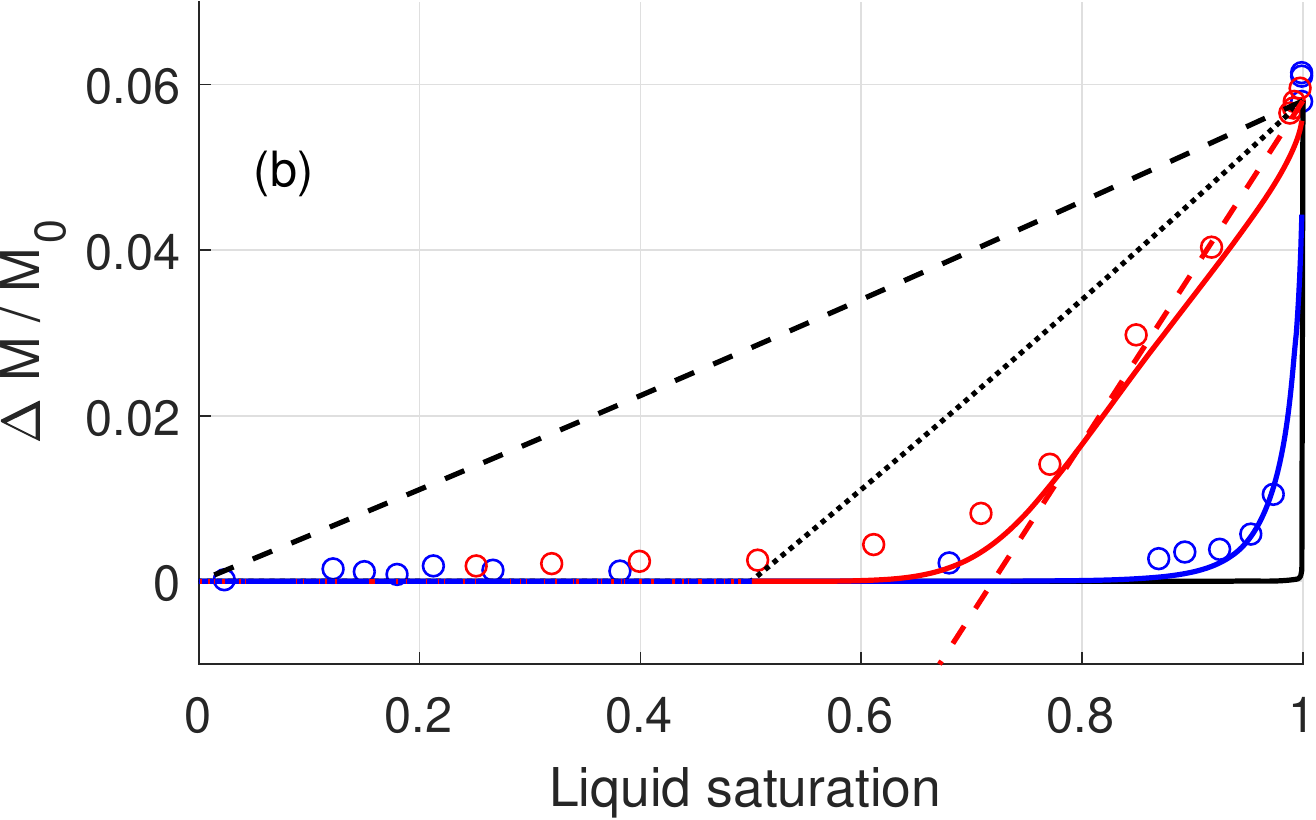}}
\caption{Relative change in the longitudinal modulus versus liquid mass fraction (saturation) during adsorption (blue) and desorption (red) of n-hexane \cite{Page1995} (a) and argon \cite{Schappert2014} (b): ultrasonic measurements (squares), best fit of the spherical shell model (solid lines), GW limit (black solid line), GH limit (black dashed line), modified GH limit (black dotted line), constant velocity (red dashed line), finite element simulations (red dotted line). }
\label{fig:hexane_modulus}
\end{figure}

If the pores are instead filled with a mixture of two fluids 1 and 2, then
the bulk modulus is defined not just by their bulk moduli $K_{f1}$ and $K_{f2},$ and volume fractions $S_{1}$ and $S_{2}=1-S_{1},$ but also by their geometrical distribution. If the two fluids are distributed uniformly within the pore space so that the pressure in the two fluids is equilibrated, then the medium can be considered as saturated with a single fluid, whose bulk modulus $K_{f}$ is given by the harmonic average of the fluid moduli \cite{Domenico:1976,Dutta:Ode:1979a,Dutta:Ode:1979b,Johnson:2001} 
\begin{equation}
\frac{1}{K_{W}}=\frac{S_{1}}{K_{f1}}+\frac{S_{2}}{K_{f2}}.  \label{Wood}
\end{equation}
Equation \ref{Wood} is known as the Wood equation, and the combination of equations \ref{Gassmann} and \ref{Wood} for the bulk modulus of
a medium saturated with a uniform (fine-scale) mixture of the two fluids, is known as the Gassmann-Wood (GW) limit $K_{GW}=K_{G}(K_{W})$ \cite{Johnson:2001,Toms:etal:2007}. However, unlike a mixture of free fluids, pressure equilibration between fluid in pores is not instant and is controlled by the permeability $\kappa$ of the porous matrix and charasteristic fluid viscosity $\eta$. According to the Biot's theory of poroelasticity \cite{Biot1956i}, fluid pressure will have enough time to equilibrate within one period of the wave with frequency $\omega $ if the characteristic size $d$ of the patches of the medium saturated with different fluids is smaller than the hydraulic diffusion length $\delta =\left( \frac{\kappa K_{f}}{\eta \phi \omega }\right) ^{1/2}$ \cite{Johnson:2001,Toms:etal:2007}. Conversely, if the patches saturated with two fluids are much larger than $\delta$, $d\gg \delta$, then the fluid pressure has no time to equilibrate between the two fluids, and hence fluid communication between these clusters can be neglected. In this case the bulk moduli $K_{G}^{(1)}=K_{G}(K_{f1})$ and $K_{G}^{(2)}=K_{G}(K_{f2})$ of the clusters saturated with fluids 1 an 2 are given by Gassmann equation \ref{Gassmann} with $K_{f}=K_{f1}$ and $K_{f}=K_{f2}$, respectively. Furthermore, Gassmann theory shows that the shear modulus of a porous medium is independent of the saturating fluids and equals to the shear modulus $G_0$ of the empty porous matrix. Thus these clusters have the same
shear modulus, and hence according to Hill's \cite{HILL1963} theorem, the bulk modulus of their mixture is uniquely defined by their volume fractions 
\begin{equation}
\label{Hill}
\frac{1}{K_{GH}+\frac{4}{3}G_0}=\frac{S_{1}}{K_{G}^{(1)} +\frac{4}{3}G_0}+\frac{S_{2}}{K_{G}^{(2)}+\frac{4}{3}G_0}. 
\end{equation}
Equation \ref{Hill} with $K_{G}^{(1)}$ and $K_{G}^{(2)}$ given by Gassmann equation is known as the Gassmann-Hill (GH) limit \cite{Norris:1993,Johnson:2001}. If the compressibilities of the two fluids are similar, then the GW and GH limits are close. However if the compressibilities are very different (say $K_{f2}\ll K_{f1}$), then the GW and GH limits are also very different. Indeed in case $K_{f2}\ll K_{f1}\ll K_{s},$ the GH limit is nearly linear in saturations
\begin{equation}
K_{GH} \simeq K_{0}+\frac{\alpha^2 }{\phi }\left( S_{1}K_{f1}+S_{2}K_{f2}\right) .
\end{equation}
Conversely $K_{GW}$ is almost independent of saturation, $K_{GW}=K_{G}^{(2)}$ until $S_{1}$ becomes close to $=1-K_{f2}/K_{f1},$ when it rises sharply to $K_{GW}=K_{G}^{(1)}$. Figure \ref{fig:hexane_modulus} shows the dependence of the relative deviation of the measured longitudinal modulus from its value at zero vapor pressure $\Delta M/M_0=(M-M_0)/M_0$ along with 
GW and GH limits for liquid and vapor adsorbates as fluids 1 and 2, respectively, calculated from the Gassmann equation at the full saturation \cite{Gor:Gurevich:2017}. Since the vapor bulk modulus is negligibly small, $K_{GW}=K_{G}^{(2)}=K_{0}$ effectively for all measurable saturations below the capillary condensation. The modulus versus saturation data show that the saturation on adsorption is closer to the GW limit and indicating more uniform saturation that on desorption. Yet the data deviates from the GW limit close to full saturation; this shows that even on adsorption, the saturation is not perfectly uniform. Hence, it is potentially possible to estimate the spatial scale of the saturation heterogeneity using dynamic patchy saturation models, which quantify the transition from GW to GH limits as patch size (or frequency) increases.

\begin{figure}[ht!]
\centerline{\includegraphics[width=0.7\linewidth]{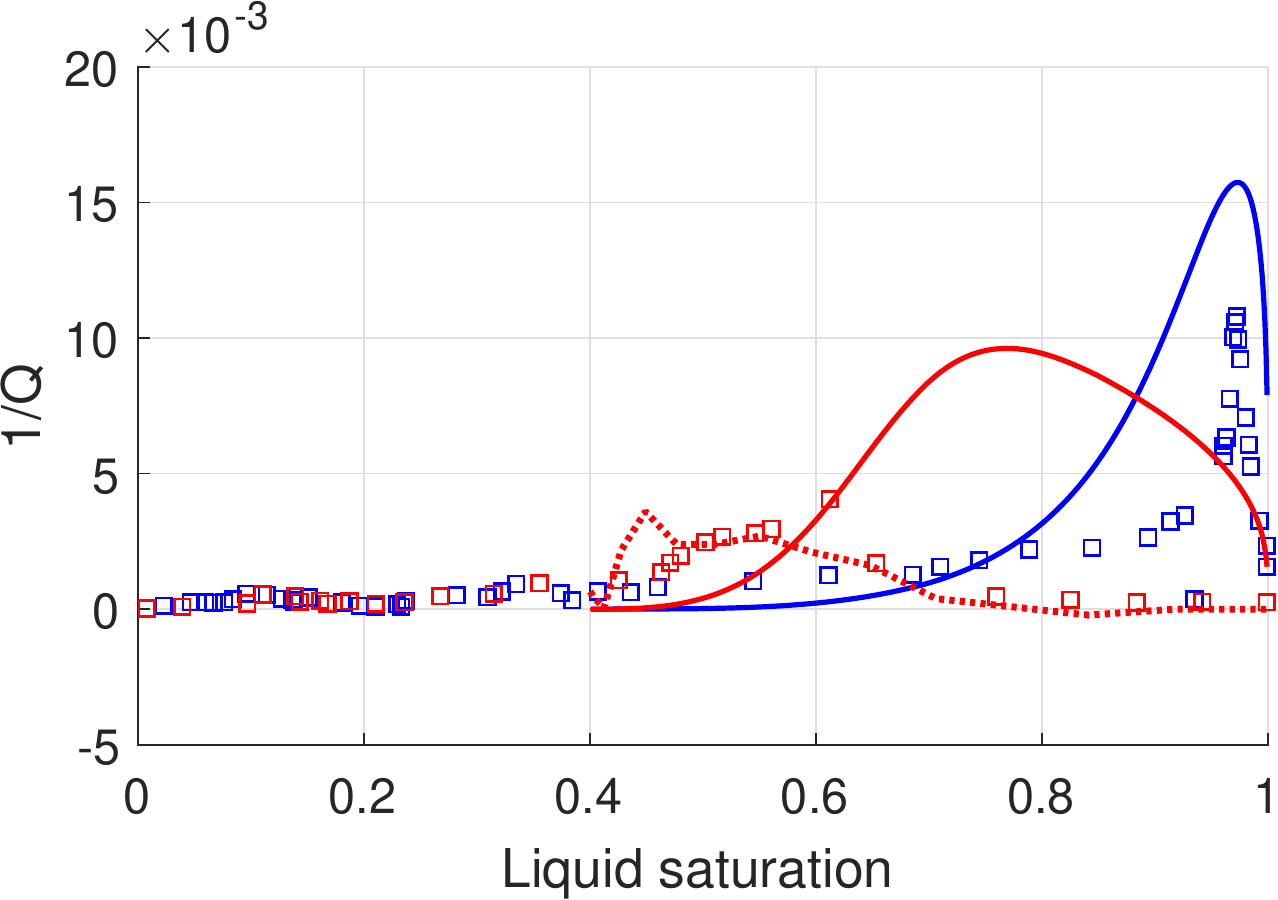}}
\caption{Longitudinal attenuation factor $Q^{-1}$ versus liquid mass fraction (saturation) during adsorption (blue) and desorption (red) of n-hexane \cite{Page1995}: ulrasonic measurements (squares), best fit of the concentric sphere model (solid lines), finite element simulations (red dotted line).}
\label{fig:hexane_Qp}
\end{figure}

\begin{figure}[ht!]
\centerline{\includegraphics[width=0.7\linewidth]{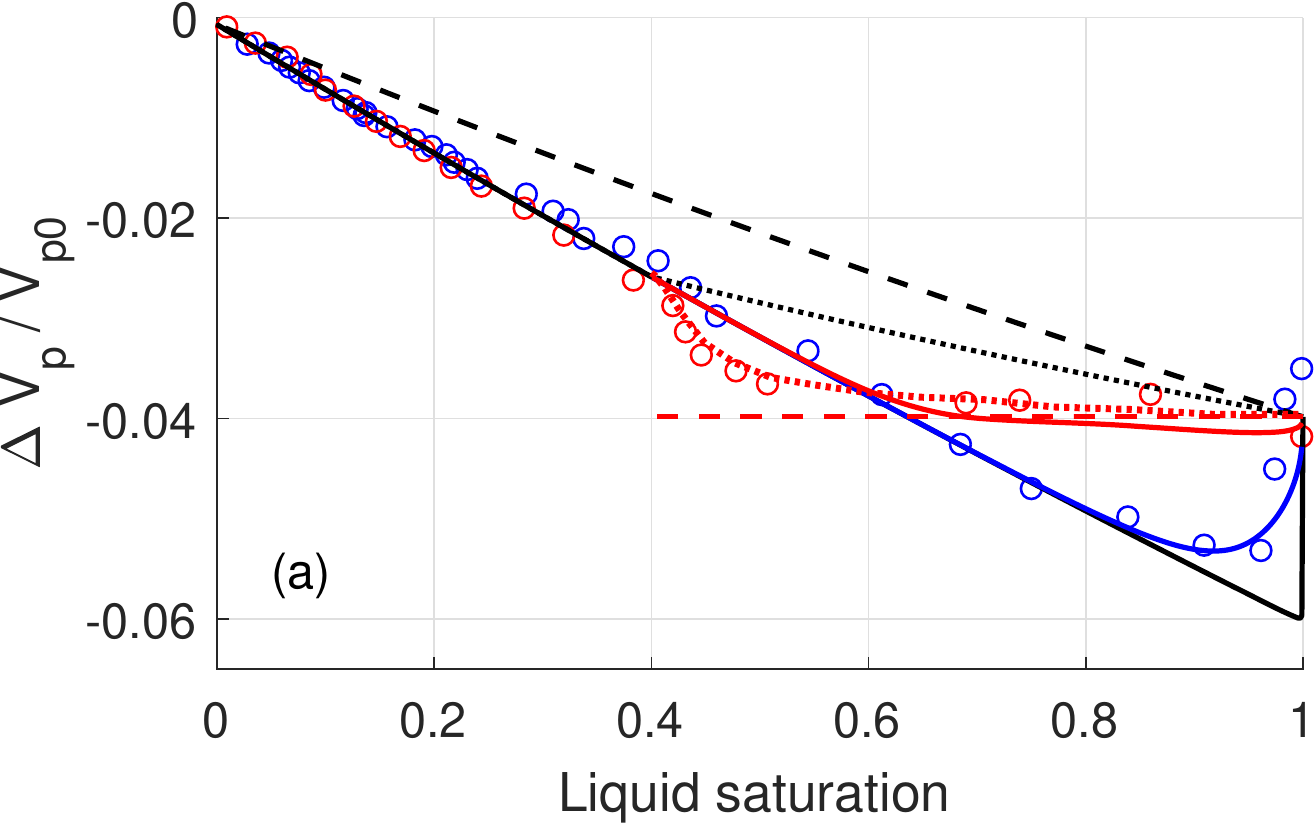}} \centerline{\includegraphics[width=0.7\linewidth]{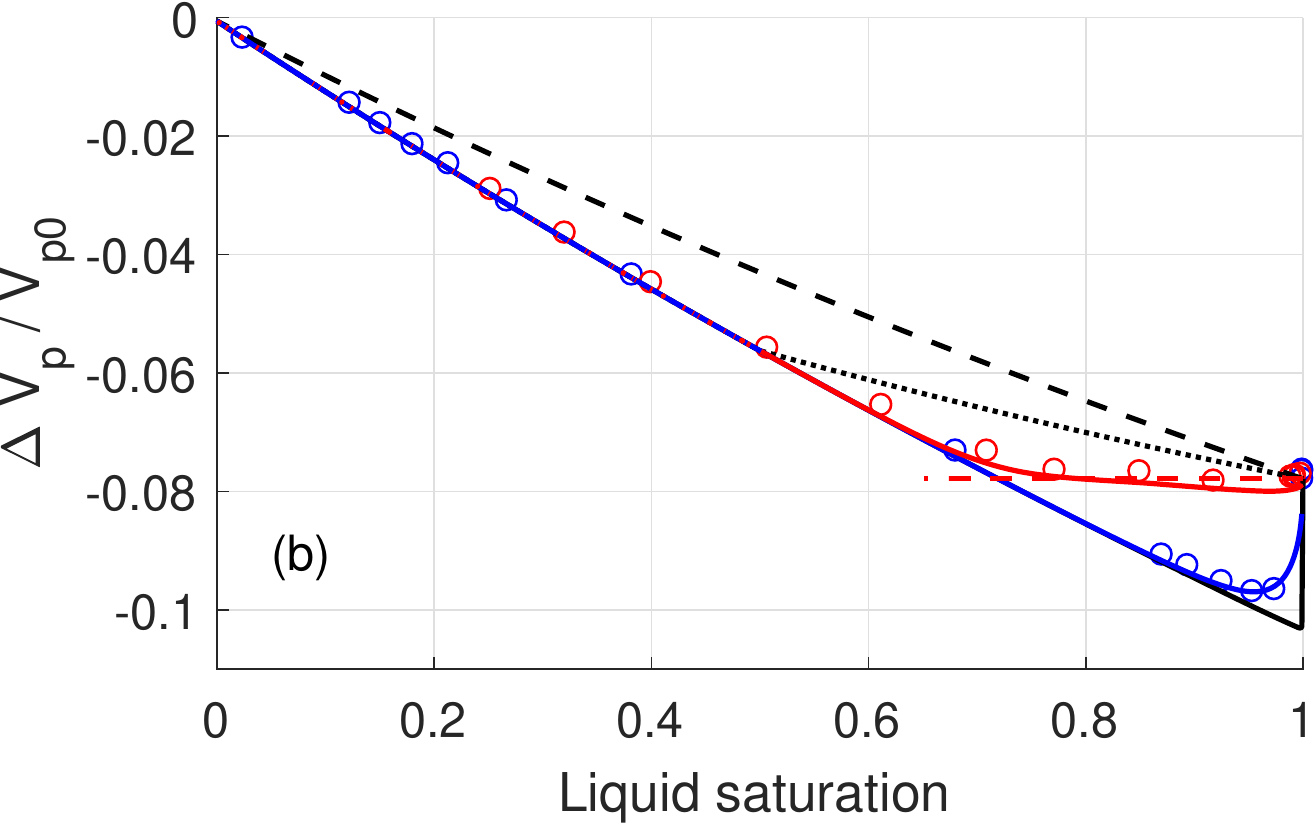}}
\caption{Relative change in the longitudinal velocity versus liquid mass fraction (saturation) during adsorption (blue) and desorption (red) of n-hexane \cite{Page1995} (a) and argon \cite{Schappert2014} (b). Notation the same as in Figure \ref{fig:hexane_modulus}. }
\label{fig:hexane_Vp}
\end{figure}

This transition depends not just on spatial scale but also on the geometry of the fluid distribution. The simplest of such models is the spherical shell model (SSM) \cite{White:1975,Dutta:Ode:1979a,Dutta:Ode:1979b,Johnson:2001}, in which the medium is assumed to consist of double spheres. Each inner sphere of radius $R_{i}$ is saturated with fluid 2 and is surrounded by an outer sphere with radius $R_{o}$, with the region between the two spheres saturated with fluid 1, so that $S_{2}=\left( R_{i}/R_{o}\right) ^{3}$ (or vice versa). A compact approximate analytical solution for the ultrasonic bulk modulus and attenuation corresponding to such a model is given in \cite[Eqs. 43-45, 40 and 34]{Johnson:2001} and briefly summarized in Supplementary Material (SM). To describe the sorption data with this model it is necessary to make some modeling choices. In particular, at non-zero vapor pressures all parts of the sample will have some liquid film on the pore walls; hence it is reasonable to assume that the saturation is always uniform below certain minimum value of liquid saturation $S_{l0}.$ The value  $S_{l0}$ depends on the specific glass sample as well as the properties of the adsorbate. For n-hexane data \cite{Page1995}, one can choose the value $S_{l0}=0.5$ corresponding to $p/p_{0}=0.4,$ the value below which the
adsorption and desorption isotherms overlap. Yet the ultrasonic moduli on adsorption and desorption only overlap below liquid saturation $S_{l}=0.46,$ indicating that ultrasonic data is more sensitive to details of fluid distribiution than mass isotherms, and suggesting that the value $S_{l0}=0.4$ is more appropriate. The same approach suggests $S_{l0}=0.5$ for the argon data of \cite{Schappert2014}.

To take into account minimum saturation of the liquid phase, we assume that fluid 1 is liquid with the modulus $K_{l}$ and saturation $S_{1}=(S_{l}-S_{l0})(1-S_{l0})$ while fluid 2 is a mixture of the liquid and vapor with the modulus $\frac{1}{K_{2}}=\frac{S_{l0}}{K_{l}}+\frac{1-S_{l0}}{K_{v}}$ and saturation $S_{2}=S_{v}/(1-S_{l0}).$ For these ``new'' fluids, the GW limit is the same as for liquid and vapor, while the new GH\ limit is shown in Fig.\ \ref{fig:hexane_modulus} by the black dotted line.

The main parameter controlling the predictions of the SSM is the radius of the outer spherical shell $R_{o}$. For \emph{adsorption} we assume that $R_{o}$ is constant (independent of saturation), which implies that the volume of the vapor ``pockets'' scales with vapor saturation ($R_{i}=R_{o}S_{2}^{1/3}$) while the number density (and distance between centres) of such pockets is constant. This assumption seems reasonable for $S_{2}<0.1$ but may break down at larger vapor saturations. On adsorption,
the model gives the best match with the saturation dependence of the modulus
for about $R_{o}=175$ nm for n-hexane and $R_{o}=130$ nm for argon.  At $S_2=0.05$ this corresponds to the size of ``vapor'' pockets of around $R_{i}=65$ nm for n-hexane and $R_{i}=48$ nm for argon. For n-hexane, \cite{Page1995} also report change of longitudinal attenuation versus saturation, and indeed the value $R_{o}=175$ nm yields a reasonable qualitative agreement with the position of the attenuation peak at $S_{l} \sim 0.95$ (Fig.\ \ref{fig:hexane_Qp}). 

For \emph{desorption} of n-hexane, the modulus shows nearly linear
dependence on saturation from full saturation down to $S_{l}=0.6$. The SSM fits these data for $R_{o} \sim 700$ nm  (red solid line in Fig.\ \ref{fig:hexane_modulus}) but predicts a large and broad attenuation peak between $S_{l}$ values of 0.6 to 0.9, which is not supported by the data (Fig.\ \ref{fig:hexane_Qp}). The linear saturation dependence might also result from saturation forming very large patches ($d\gg \delta $) from full saturation down to $S_{l}=0.6$, below which it becomes uniform. However such behavior does not explain the attenuation peak around $S_{l}=0.5$; it is also unreasonable to assume that saturation on desorption is more uniform than on adsorption. 

A simpler and much more convincing explanation can be inferred from the behavior of compressional wave velocity as a function of saturation, as plotted in Fig.\ \ref{fig:hexane_Vp}. n-hexane desorption data show that the velocity remains nearly constant from full saturation down to $S_{l} \sim 0.5$. This is consistent with the optical observations \cite{Page1995} that ``drying'' of the cylindrical sample begins from the surface, while the middle core remains fully saturated. As relative pressure and liquid saturation are reduced, the wave travels through this fully saturated core with the same velocity and is not affected by the reduced overall saturation until this core becomes relatively thin, and then recorded arrival switches to the unsaturated (``dried'') outer shell of the cylinder. This explains the reduction of the modulus (Fig.\ \ref{fig:hexane_modulus}) below that in the dry sample: for $S_{l}>0.5$ the wave travels through the saturated region with the velocity $v_{l}=\sqrt{M_{G}(K_{l})/\rho _{sat}},$ where $M_{G}(K_{l})=K_{G}(K_{l})+(4/3)G_0$ is the longitudinal modulus and $\rho _{sat}=\rho _{0}+\phi \rho _{l}$ is the density of the fully saturated sample (where $\rho _{0}$ and $\rho_{l}$ are the densities of the dry porous glass and liquid adsorbate, respectively). However, the apparent modulus in Fig.\ \ref{fig:hexane_modulus} is computed as $M=\rho_{b}v_{l}^{2}$, where $\rho _{b}=\rho _{0}+\phi S_{l}\rho _{l}=\rho_{sat}-\phi (1-S_{l})\rho _{l}$ (where we neglected the term with vapor density). Hence 
\begin{equation}
\label{modulusD}
\frac{\Delta M}{M_{0}}=\frac{\rho _{b}M_{G}(K_{l})}{\rho _{sat}M_{0}}-1=\frac{1+\frac{\alpha K_{l}}{\phi K_{0}}}{1+\frac{\phi \rho _{l}}{\rho _{0}}} \simeq \frac{\alpha K_{l}}{\phi K_{0}}-\frac{\phi \rho _{l}}{\rho_{0}}(1-S_{l}).  
\end{equation}
Thus at full liquid saturation $\Delta M/M_{0}$ is positive but decreases linearly as $S_{l}$ decreases, becoming negative at $S_{l}=0.5$. We note that for macroscopically heterogeneous materials (with the heterogeneity scale $d$ larger than the wavelength), calculation of the modulus from velocities is somewhat unphysical, as the wave speeds are controlled by distribution of densities as well as moduli, whereas the true effective modulus (ratio of stress to strain) depends on the distribution of moduli only. Our interpretation also
explains the apparent attenuation peak around $S_{l}=0.5$. Indeed at this saturation the ultrasonic energy is split between two arrivals (traveling through the saturated and dry portions of the sample) and hence their amplitudes are lower than in the fully saturated, dry or uniformly saturated sample. This interpretation of the ultrasonic moduli and \textit{apparent} attenuation on desorption of n-Hexane is supported by the results of finite element simulations shown as red dotted lines in Figs \ref{fig:hexane_modulus}a, \ref{fig:hexane_Qp} and \ref{fig:hexane_Vp}a, and detailed in SM. Argon desorption data (Figs. \ref{fig:hexane_modulus}b, \ref{fig:hexane_Vp}b) show similar behavior, but without the anomalously low apparent modulus (Fig.\ \ref{fig:hexane_modulus}).  This may be a result of higher sensitivity of ultrasonic transducers (capable of detecting weak early arrivals) or a different algorithm for picking arrival times.

In summary, we have performed a poroelastic analysis of the dependence of ultrasonic moduli of Vycor glass on vapor pressure as measured during sorption experiments. This analysis shows that both on adsorption and desorption of argon and n-hexane, the condensate in the pore space forms patches much larger than the typical pore radius. The patch sizes are much larger on desorption than on adsorption. On adsorption the patch diameter is on the order of 10-20 pore diameters, while on desorption the patch size is comparable to the sample size.

These results suggest that ultrasonic measurements are a promising method for studying fluid distributions during sorption. More ultrasonic measurements on different porous materials with different adsorbates are required to better understand the fluid distributions in these processes. 

\vspace{0.5cm}

\noindent G.G. thanks Patrick Huber for pointing out some of the references discussed in this work. B.G. thanks the sponsors of the Curtin Reservoir Geophysics Consortium for financial support, and Julianna Toms and Eva Caspari for discussions of implementation of the methods in Refs. \onlinecite{Toms:etal:2007, Toms2008}.


\begin{thebibliography}{31}
\expandafter\ifx\csname natexlab\endcsname\relax\def\natexlab#1{#1}\fi
\expandafter\ifx\csname bibnamefont\endcsname\relax
  \def\bibnamefont#1{#1}\fi
\expandafter\ifx\csname bibfnamefont\endcsname\relax
  \def\bibfnamefont#1{#1}\fi
\expandafter\ifx\csname citenamefont\endcsname\relax
  \def\citenamefont#1{#1}\fi
\expandafter\ifx\csname url\endcsname\relax
  \def\url#1{\texttt{#1}}\fi
\expandafter\ifx\csname urlprefix\endcsname\relax\def\urlprefix{URL }\fi
\providecommand{\bibinfo}[2]{#2}
\providecommand{\eprint}[2][]{\url{#2}}

\bibitem[{\citenamefont{Rouquerol et~al.}(2013)\citenamefont{Rouquerol,
  Rouquerol, Llewellyn, Maurin, and Sing}}]{Rouquerol2013}
\bibinfo{author}{\bibfnamefont{J.}~\bibnamefont{Rouquerol}},
  \bibinfo{author}{\bibfnamefont{F.}~\bibnamefont{Rouquerol}},
  \bibinfo{author}{\bibfnamefont{P.}~\bibnamefont{Llewellyn}},
  \bibinfo{author}{\bibfnamefont{G.}~\bibnamefont{Maurin}}, \bibnamefont{and}
  \bibinfo{author}{\bibfnamefont{K.~S.} \bibnamefont{Sing}},
  \emph{\bibinfo{title}{Adsorption by powders and porous solids: principles,
  methodology and applications}} (\bibinfo{publisher}{Academic press},
  \bibinfo{year}{2013}).

\bibitem[{\citenamefont{Gor et~al.}(2017)\citenamefont{Gor, Huber, and
  Bernstein}}]{Gor2017review}
\bibinfo{author}{\bibfnamefont{G.~Y.} \bibnamefont{Gor}},
  \bibinfo{author}{\bibfnamefont{P.}~\bibnamefont{Huber}}, \bibnamefont{and}
  \bibinfo{author}{\bibfnamefont{N.}~\bibnamefont{Bernstein}},
  \bibinfo{journal}{Appl. Phys. Rev.} \textbf{\bibinfo{volume}{4}},
  \bibinfo{pages}{011303} (\bibinfo{year}{2017}).

\bibitem[{\citenamefont{Pan and Connell}(2007)}]{Pan2007}
\bibinfo{author}{\bibfnamefont{Z.}~\bibnamefont{Pan}} \bibnamefont{and}
  \bibinfo{author}{\bibfnamefont{L.~D.} \bibnamefont{Connell}},
  \bibinfo{journal}{Int. J. Coal Geol.} \textbf{\bibinfo{volume}{69}},
  \bibinfo{pages}{243} (\bibinfo{year}{2007}).

\bibitem[{\citenamefont{Barthelemy et~al.}(2007)\citenamefont{Barthelemy,
  Ghulinyan, Gaburro, Toninelli, Pavesi, and Wiersma}}]{Barthelemy2007}
\bibinfo{author}{\bibfnamefont{P.}~\bibnamefont{Barthelemy}},
  \bibinfo{author}{\bibfnamefont{M.}~\bibnamefont{Ghulinyan}},
  \bibinfo{author}{\bibfnamefont{Z.}~\bibnamefont{Gaburro}},
  \bibinfo{author}{\bibfnamefont{C.}~\bibnamefont{Toninelli}},
  \bibinfo{author}{\bibfnamefont{L.}~\bibnamefont{Pavesi}}, \bibnamefont{and}
  \bibinfo{author}{\bibfnamefont{D.~S.} \bibnamefont{Wiersma}},
  \bibinfo{journal}{Nat. Photonics} \textbf{\bibinfo{volume}{1}},
  \bibinfo{pages}{172} (\bibinfo{year}{2007}).

\bibitem[{\citenamefont{Varanakkottu et~al.}(2014)\citenamefont{Varanakkottu,
  Engelbart, Joshi, Still, Xiao, and Hardt}}]{Varanakkottu2014}
\bibinfo{author}{\bibfnamefont{S.~N.} \bibnamefont{Varanakkottu}},
  \bibinfo{author}{\bibfnamefont{H.}~\bibnamefont{Engelbart}},
  \bibinfo{author}{\bibfnamefont{S.}~\bibnamefont{Joshi}},
  \bibinfo{author}{\bibfnamefont{M.}~\bibnamefont{Still}},
  \bibinfo{author}{\bibfnamefont{W.}~\bibnamefont{Xiao}}, \bibnamefont{and}
  \bibinfo{author}{\bibfnamefont{S.}~\bibnamefont{Hardt}},
  \bibinfo{journal}{Opt. Express} \textbf{\bibinfo{volume}{22}},
  \bibinfo{pages}{25560} (\bibinfo{year}{2014}).

\bibitem[{\citenamefont{Page et~al.}(1995)\citenamefont{Page, Liu, Abeles,
  Herbolzheimer, Deckman, and Weitz}}]{Page1995}
\bibinfo{author}{\bibfnamefont{J.~H.} \bibnamefont{Page}},
  \bibinfo{author}{\bibfnamefont{J.}~\bibnamefont{Liu}},
  \bibinfo{author}{\bibfnamefont{B.}~\bibnamefont{Abeles}},
  \bibinfo{author}{\bibfnamefont{E.}~\bibnamefont{Herbolzheimer}},
  \bibinfo{author}{\bibfnamefont{H.~W.} \bibnamefont{Deckman}},
  \bibnamefont{and} \bibinfo{author}{\bibfnamefont{D.~A.} \bibnamefont{Weitz}},
  \bibinfo{journal}{Phys. Rev. E} \textbf{\bibinfo{volume}{52}},
  \bibinfo{pages}{2763} (\bibinfo{year}{1995}).

\bibitem[{\citenamefont{Toms et~al.}(2007)\citenamefont{Toms, M\"{u}ller, and
  Gurevich}}]{Toms:etal:2007}
\bibinfo{author}{\bibfnamefont{J.}~\bibnamefont{Toms}},
  \bibinfo{author}{\bibfnamefont{T.~M.} \bibnamefont{M\"{u}ller}},
  \bibnamefont{and} \bibinfo{author}{\bibfnamefont{B.}~\bibnamefont{Gurevich}},
  \bibinfo{journal}{Geophys. Prospect.} \textbf{\bibinfo{volume}{55}},
  \bibinfo{pages}{671} (\bibinfo{year}{2007}), ISSN \bibinfo{issn}{1365-2478}.

\bibitem[{\citenamefont{Schappert and Pelster}(2014)}]{Schappert2014}
\bibinfo{author}{\bibfnamefont{K.}~\bibnamefont{Schappert}} \bibnamefont{and}
  \bibinfo{author}{\bibfnamefont{R.}~\bibnamefont{Pelster}},
  \bibinfo{journal}{Europhys. Lett.} \textbf{\bibinfo{volume}{105}},
  \bibinfo{pages}{56001} (\bibinfo{year}{2014}).

\bibitem[{\citenamefont{Domenico}(1976)}]{Domenico:1976}
\bibinfo{author}{\bibfnamefont{S.~N.} \bibnamefont{Domenico}},
  \bibinfo{journal}{Geophysics} \textbf{\bibinfo{volume}{41}},
  \bibinfo{pages}{882} (\bibinfo{year}{1976}).

\bibitem[{\citenamefont{Dutta and
  Od\'{e}}(1979{\natexlab{a}})}]{Dutta:Ode:1979a}
\bibinfo{author}{\bibfnamefont{N.}~\bibnamefont{Dutta}} \bibnamefont{and}
  \bibinfo{author}{\bibfnamefont{H.}~\bibnamefont{Od\'{e}}},
  \bibinfo{journal}{Geophysics} \textbf{\bibinfo{volume}{44}},
  \bibinfo{pages}{1777} (\bibinfo{year}{1979}{\natexlab{a}}).

\bibitem[{\citenamefont{Dutta and
  Od\'{e}}(1979{\natexlab{b}})}]{Dutta:Ode:1979b}
\bibinfo{author}{\bibfnamefont{N.}~\bibnamefont{Dutta}} \bibnamefont{and}
  \bibinfo{author}{\bibfnamefont{H.}~\bibnamefont{Od\'{e}}},
  \bibinfo{journal}{Geophysics} \textbf{\bibinfo{volume}{44}},
  \bibinfo{pages}{1789} (\bibinfo{year}{1979}{\natexlab{b}}).

\bibitem[{\citenamefont{Murphy}(1982)}]{Murphy1982}
\bibinfo{author}{\bibfnamefont{W.~F.} \bibnamefont{Murphy}},
  \bibinfo{journal}{J. Acoust. Soc. Am.} \textbf{\bibinfo{volume}{71}},
  \bibinfo{pages}{1458} (\bibinfo{year}{1982}).

\bibitem[{\citenamefont{Cadoret et~al.}(1998)\citenamefont{Cadoret, Mavko, and
  Zinszner}}]{Cadoret:etal:1998}
\bibinfo{author}{\bibfnamefont{T.}~\bibnamefont{Cadoret}},
  \bibinfo{author}{\bibfnamefont{G.}~\bibnamefont{Mavko}}, \bibnamefont{and}
  \bibinfo{author}{\bibfnamefont{B.}~\bibnamefont{Zinszner}},
  \bibinfo{journal}{Geophysics} \textbf{\bibinfo{volume}{63}},
  \bibinfo{pages}{154} (\bibinfo{year}{1998}).

\bibitem[{\citenamefont{Knight et~al.}(1998)\citenamefont{Knight, Dvorkin, and
  Nur}}]{Knight:Dvorkin:Nur:1998}
\bibinfo{author}{\bibfnamefont{R.}~\bibnamefont{Knight}},
  \bibinfo{author}{\bibfnamefont{J.}~\bibnamefont{Dvorkin}}, \bibnamefont{and}
  \bibinfo{author}{\bibfnamefont{A.}~\bibnamefont{Nur}},
  \bibinfo{journal}{Geophysics} \textbf{\bibinfo{volume}{63}},
  \bibinfo{pages}{132} (\bibinfo{year}{1998}).

\bibitem[{\citenamefont{Johnson}(2001)}]{Johnson:2001}
\bibinfo{author}{\bibfnamefont{D.~L.} \bibnamefont{Johnson}},
  \bibinfo{journal}{J. Acoust. Soc. Am.} \textbf{\bibinfo{volume}{110}},
  \bibinfo{pages}{682} (\bibinfo{year}{2001}).

\bibitem[{\citenamefont{Toms}(2008)}]{Toms2008}
\bibinfo{author}{\bibfnamefont{J.~J.} \bibnamefont{Toms}}, Ph.D. thesis,
  \bibinfo{school}{Curtin University of Technology} (\bibinfo{year}{2008}).

\bibitem[{\citenamefont{Caspari et~al.}(2011)\citenamefont{Caspari, M\"uller,
  and Gurevich}}]{Caspari:Muller:Gurevich:2011}
\bibinfo{author}{\bibfnamefont{E.}~\bibnamefont{Caspari}},
  \bibinfo{author}{\bibfnamefont{T.~M.} \bibnamefont{M\"uller}},
  \bibnamefont{and} \bibinfo{author}{\bibfnamefont{B.}~\bibnamefont{Gurevich}},
  \bibinfo{journal}{Geophys. Res. Lett.} \textbf{\bibinfo{volume}{38}},
  \bibinfo{pages}{L13301} (\bibinfo{year}{2011}).

\bibitem[{\citenamefont{Rubino and Holliger}(2012)}]{Rubino:Holliger:2012}
\bibinfo{author}{\bibfnamefont{J.~G.} \bibnamefont{Rubino}} \bibnamefont{and}
  \bibinfo{author}{\bibfnamefont{K.}~\bibnamefont{Holliger}},
  \bibinfo{journal}{Geophysical Journal International}
  \textbf{\bibinfo{volume}{188}}, \bibinfo{pages}{1088} (\bibinfo{year}{2012}).

\bibitem[{\citenamefont{Mavko and Nur}(1975)}]{Mavko:Nur:1975}
\bibinfo{author}{\bibfnamefont{G.}~\bibnamefont{Mavko}} \bibnamefont{and}
  \bibinfo{author}{\bibfnamefont{A.}~\bibnamefont{Nur}}, \bibinfo{journal}{J.
  Geophys. Res.} \textbf{\bibinfo{volume}{80}}, \bibinfo{pages}{1444}
  (\bibinfo{year}{1975}).

\bibitem[{\citenamefont{Jones}(1986)}]{Jones:1986}
\bibinfo{author}{\bibfnamefont{T.}~\bibnamefont{Jones}},
  \bibinfo{journal}{Geophysics} \textbf{\bibinfo{volume}{51}},
  \bibinfo{pages}{1939} (\bibinfo{year}{1986}).

\bibitem[{\citenamefont{Murphy~III et~al.}(1986)\citenamefont{Murphy~III,
  Winkler, and Kleinberg}}]{Murphy:etal:1986}
\bibinfo{author}{\bibfnamefont{W.~F.} \bibnamefont{Murphy~III}},
  \bibinfo{author}{\bibfnamefont{K.~W.} \bibnamefont{Winkler}},
  \bibnamefont{and} \bibinfo{author}{\bibfnamefont{R.~L.}
  \bibnamefont{Kleinberg}}, \bibinfo{journal}{Geophysics}
  \textbf{\bibinfo{volume}{51}}, \bibinfo{pages}{757} (\bibinfo{year}{1986}).

\bibitem[{\citenamefont{Mavko and Jizba}(1991)}]{Mavko1991}
\bibinfo{author}{\bibfnamefont{G.}~\bibnamefont{Mavko}} \bibnamefont{and}
  \bibinfo{author}{\bibfnamefont{D.}~\bibnamefont{Jizba}},
  \bibinfo{journal}{Geophysics} \textbf{\bibinfo{volume}{56}},
  \bibinfo{pages}{1940} (\bibinfo{year}{1991}).

\bibitem[{\citenamefont{Gurevich et~al.}(2010)\citenamefont{Gurevich,
  Makarynska, de~Paula, and Pervukhina}}]{Gurevich2010}
\bibinfo{author}{\bibfnamefont{B.}~\bibnamefont{Gurevich}},
  \bibinfo{author}{\bibfnamefont{D.}~\bibnamefont{Makarynska}},
  \bibinfo{author}{\bibfnamefont{O.~B.} \bibnamefont{de~Paula}},
  \bibnamefont{and}
  \bibinfo{author}{\bibfnamefont{M.}~\bibnamefont{Pervukhina}},
  \bibinfo{journal}{Geophysics} \textbf{\bibinfo{volume}{75}},
  \bibinfo{pages}{N109} (\bibinfo{year}{2010}).

\bibitem[{\citenamefont{M{\"u}ller et~al.}(2010)\citenamefont{M{\"u}ller,
  Gurevich, and Lebedev}}]{Muller:Gurevich:Lebedev:2010}
\bibinfo{author}{\bibfnamefont{T.~M.} \bibnamefont{M{\"u}ller}},
  \bibinfo{author}{\bibfnamefont{B.}~\bibnamefont{Gurevich}}, \bibnamefont{and}
  \bibinfo{author}{\bibfnamefont{M.}~\bibnamefont{Lebedev}},
  \bibinfo{journal}{Geophysics} \textbf{\bibinfo{volume}{75}},
  \bibinfo{pages}{75A147} (\bibinfo{year}{2010}).

\bibitem[{\citenamefont{Gassmann}(1951)}]{Gassmann1951}
\bibinfo{author}{\bibfnamefont{F.}~\bibnamefont{Gassmann}},
  \bibinfo{journal}{Viertel. Naturforsch. Ges. Z\"urich}
  \textbf{\bibinfo{volume}{96}}, \bibinfo{pages}{1} (\bibinfo{year}{1951}).

\bibitem[{\citenamefont{Berryman}(1999)}]{Berryman1999}
\bibinfo{author}{\bibfnamefont{J.~G.} \bibnamefont{Berryman}},
  \bibinfo{journal}{Geophysics} \textbf{\bibinfo{volume}{64}},
  \bibinfo{pages}{1627} (\bibinfo{year}{1999}).

\bibitem[{\citenamefont{Gor and Gurevich}(2018)}]{Gor:Gurevich:2017}
\bibinfo{author}{\bibfnamefont{G.~Y.} \bibnamefont{Gor}} \bibnamefont{and}
  \bibinfo{author}{\bibfnamefont{B.}~\bibnamefont{Gurevich}},
  \bibinfo{journal}{Geophys. Res. Lett.} \textbf{\bibinfo{volume}{45}},
  \bibinfo{pages}{146} (\bibinfo{year}{2018}).

\bibitem[{\citenamefont{Biot}(1956)}]{Biot1956i}
\bibinfo{author}{\bibfnamefont{M.~A.} \bibnamefont{Biot}}, \bibinfo{journal}{J.
  Acoust. Soc. Am.} \textbf{\bibinfo{volume}{28}}, \bibinfo{pages}{168}
  (\bibinfo{year}{1956}).

\bibitem[{\citenamefont{Hill}(1963)}]{HILL1963}
\bibinfo{author}{\bibfnamefont{R.}~\bibnamefont{Hill}}, \bibinfo{journal}{J.
  Mech. Phys. Solids} \textbf{\bibinfo{volume}{11}}, \bibinfo{pages}{357 }
  (\bibinfo{year}{1963}), ISSN \bibinfo{issn}{0022-5096}.

\bibitem[{\citenamefont{Norris}(1993)}]{Norris:1993}
\bibinfo{author}{\bibfnamefont{A.~N.} \bibnamefont{Norris}},
  \bibinfo{journal}{J. Acoust. Soc. Am.} \textbf{\bibinfo{volume}{94}},
  \bibinfo{pages}{359} (\bibinfo{year}{1993}).

\bibitem[{\citenamefont{White}(1975)}]{White:1975}
\bibinfo{author}{\bibfnamefont{J.~E.} \bibnamefont{White}},
  \bibinfo{journal}{Geophysics} \textbf{\bibinfo{volume}{40}},
  \bibinfo{pages}{224} (\bibinfo{year}{1975}).

\end{thebibliography}

\end{document}